\documentclass[12pt, letterpaper]{article}

\usepackage{amsmath}
\usepackage{amsfonts}
\usepackage{amssymb}
\usepackage{graphicx}

\pdfoutput=1

\usepackage{color}

\usepackage{amsmath, amssymb, graphics, epsfig, graphicx}
\usepackage{epsf}
\usepackage{epstopdf}
\usepackage {amssymb}
\newcommand{\nc}{\newcommand}
\nc{\ba}{\begin{eqnarray}}
\nc{\ea}{\end{eqnarray}}
\newcommand\be{\begin{equation}}
\newcommand\ee{\end{equation}}

\newcommand{\calR}{{\cal{R}}}

\newcommand{\calP}{{\cal{P}}}

\newcommand{\bea}{\begin{eqnarray}}
\newcommand{\eea}{\end{eqnarray}}

\newcommand{\bfx}{{\bf{x}}}
\newcommand{\bfq}{{\bf{q}}}
\newcommand{\bfp}{{\bf{p}}}
\newcommand{\bfk}{{\bf{k}}}
\newcommand{\Ha}{{\bf H}_3  }

\begin{document}

\vspace{5mm}
\vspace{0.5cm}
\begin{center}

\def\thefootnote{\fnsymbol{footnote}}

{ \bf   Loop Corrections in Gravitational Wave Spectrum in Single Field Inflation }
\\[1cm]

{ 
Hassan Firouzjahi$\footnote{firouz@ipm.ir}$
}

{\small \textit{ School of Astronomy, Institute for Research in Fundamental Sciences (IPM) \\ P.~O.~Box 19395-5531, Tehran, Iran
}}\\

\end{center}

\vspace{.8cm}

\hrule \vspace{0.3cm}


\begin{abstract}

We study the one-loop corrections in power spectrum of long gravitational waves induced from small scale modes in the models of single field inflation undergoing  a phase of ultra-slow-roll (USR).  We show that the spectrum of long tensor perturbations  are largely unaffected by the loop corrections from the short scalar modes. In particular, the spectrum of  long tensor perturbations 
is insensitive to the sharpness of the transition from the USR phase to the final slow-roll phase. This is in contrast to the case of scalar power spectrum  in which the loop corrections can be large for a sharp transition  while it is slow-roll suppressed in a mild transition. 
We study  the tensor-scalar-scalar bispectrum in the squeezed limit
and  demonstrate that the Maldacena consistency condition does hold.

\end{abstract}
\vspace{0.5cm} \hrule
\def\thefootnote{\arabic{footnote}}
\setcounter{footnote}{0}
\newpage
\section{Introduction}
\label{intro}

Recently the question of one-loop corrections in the power spectrum of large CMB scale scalar perturbations from the small scale modes in the setup of single field inflation undergoing a phase of ultra-slow-roll (USR) was debated extensively \cite{Kristiano:2022maq, Kristiano:2023scm, Riotto:2023gpm, Riotto:2023hoz, Choudhury:2023vuj,  Choudhury:2023jlt, Choudhury:2023rks, 
Firouzjahi:2023aum, Motohashi:2023syh, Firouzjahi:2023ahg}, for a related earlier work see \cite{Cheng:2021lif}.  This is particularly an important question since the models of single field inflation with an intermediate USR phase  have been employed extensively in recent years as a viable mechanism to generate primordial black holes (PBHs) which may comprise all or parts of cold dark matter \cite{Ivanov:1994pa, Garcia-Bellido:2017mdw, Biagetti:2018pjj}, for a review see \cite{Ozsoy:2023ryl, Byrnes:2021jka}. More specifically, to have a successful mechanism of PBHs formation, one requires the amplitude of curvature perturbations to be enhanced by a factor of $10^7$ or so in the allowed small  scales compared to the large CMB scales. It turns out that an intermediate phase of USR inflation can provide this enhancement naturally.  

The USR setup is a phase of inflation in which the potential is very flat \cite{Kinney:2005vj, Morse:2018kda, Lin:2019fcz}. Consequently, the inflaton velocity falls off exponentially and the curvature perturbations grow on superhorizon scales \cite{Namjoo:2012aa}. As the curvature perturbations grows on superhorizon scales, it provides a non-trivial example for the violation of the celebrated Maldacena consistency condition \cite{Maldacena:2002vr, Creminelli:2004yq} for the non-Gaussianity of single field inflation \cite{Namjoo:2012aa, Martin:2012pe, Chen:2013aj, Chen:2013eea, Akhshik:2015rwa, Mooij:2015yka, Bravo:2017wyw, Finelli:2017fml, Pi:2022ysn}. More specifically, it was shown in \cite{Namjoo:2012aa} that the amplitude of local-type non-Gaussianity in USR model is $f_{NL}=\frac{5}{2}$. This question was further investigated in \cite{Cai:2018dkf} in which it was demonstrated that the final amplitude of $f_{NL}$ crucially depends on the sharpness of the transition from the USR phase to the final slow-roll (SR) phase. In particular, in an extreme sharp transition from the USR phase to the SR phase, as assumed in \cite{Namjoo:2012aa}, $f_{NL}$ reaches its maximum value $\frac{5}{2}$. However, if the transition is mild, then the curvature perturbations evolve after the USR phase until it reaches to is final attractor value. Correspondingly, much of the amplitude of $f_{NL}$ is washed out and it ends up to a value at the order of the slow-roll parameters though the  Maldacena consistency condition is still violated. The lesson is that the sharpness of the transition from the USR phase to the final SR phase plays important roles to read off  the amplitude of cosmological observables  at the end of inflation. 

Originally, it was argued in \cite{Kristiano:2022maq}, see also  \cite{Kristiano:2023scm}, that the  one-loop corrections from small USR modes  can significantly affect the large CMB scale modes. Therefore, it was argued that to keep these loop corrections under perturbative control, the model loses its applicability to generate the desired PBHs abundance. This conclusion was criticized in  \cite{ Riotto:2023gpm, Riotto:2023hoz} where it was advocated that this conclusion  is model-dependent and the dangerous one-loop corrections can be harmless in a smooth transition. This question was further investigated in \cite{Firouzjahi:2023aum} in a consistent manner where the effects of both cubic and quartic Hamiltonians were taken into account. While the analysis in \cite{Firouzjahi:2023aum} supported the conclusion of  \cite{Kristiano:2022maq} 
for the setup with a  sharp transition but it was argued that the situation can be very different in a mild transition. Finally, this question was further studied in  \cite{Firouzjahi:2023ahg}  where, using $\delta N$ formalism, it was shown that for a mild transition the one-loop corrections are suppressed by the slow-roll parameters and the setup can still be viable for PBHs formation,  in agreement with  \cite{ Riotto:2023gpm, Riotto:2023hoz}. The conclusion from these works, as in the old story of $f_{NL}$ alluded to before, is that the amplitude of one-loop corrections crucially depends on the sharpness of the transition from the USR phase to the final SR phase. For a physical smooth transition, the dangerous one-loop corrections are washed out during the subsequent evolutions of the modes after the USR phase. 

With the above discussions in mind, in this work we extend the motivation of \cite{Kristiano:2022maq} and calculate the one-loop correction from small USR modes on large CMB scale gravitational waves (GWs) perturbations \footnote{For earlier works concerning then loop corrections in tensor power spectrum during inflation see \cite{Ota:2022xni, Chen:2022dah, Ota:2022hvh, Meng:2022ixx, Brahma:2022yxu}. }. On the physical ground, similar to the reasonings of  \cite{ Riotto:2023gpm, Riotto:2023hoz},  it is expected that the tensor perturbations to be less sensitive to the USR phase transition. This is because the amplitude of GWs are determined by the Hubble scale, $H$, during inflation. As the value of $H$ is not much modified during the USR transition, then the background for GWs propagation is not much modified either.  Add to it the important effect that the tensor perturbations are frozen on superhorizon scales  at the linear level in perturbation theory \cite{Weinberg:2008zzc, Baumann:2022mni, Kodama:1984ziu, Mukhanov:1990me}. However, the lesson  of large loop corrections in a sharp transition for the case of scalar power spectrum sets a non-trivial example to examine more directly the validity of the above physical expectations for the long GWs. This is the goal of this work.


\section{The Setup}
\label{setup}
Here we briefly review our setup and present the formulas which will be required for our subsequent analysis. 

We consider a three-phase model of inflation  
in which a USR phase is sandwiched between two phases of SR inflation ($SR \rightarrow USR \rightarrow SR$). The early SR phase is when the large CMB scale mode leaves the horizon.  The USR phase is extended in the interval $t_i \leq t \leq t_e$ in which the potential is flat $V(\phi)= V_0$. The background equations during the USR phase are, 
\ba
\ddot \phi(t) + 3 H \dot \phi(t)=0\, , \quad \quad 3 M_P^2 H^2 \simeq V_0, 
\ea
where $M_P$ is the reduced Planck mass and  $H$ is the Hubble expansion rate during inflation. During the USR phase  $H$ is very nearly constant while 
$\dot \phi \propto \frac{1}{a^3}$. 

The two slow-roll  parameters related to  $H$  are given as follows,
\ba
\label{ep-eta}
\epsilon \equiv -\frac{\dot H}{H^2} =\frac{\dot \phi^2}{2 M_P^2 H^2}\, , \quad \quad 
\eta \equiv \frac{\dot \epsilon}{H \epsilon} \, . 
\ea
Since $\epsilon$ falls off like $a^{-6}$ during the USR setup, we see that $\eta \simeq -6$ which is the hallmark of the USR inflation \cite{Kinney:2005vj}.  Going to conformal time $d \tau= dt/a(t)$ with $a H \tau \simeq -1$, 
 the evolution of $\epsilon$ is given by 
 \ba
 \epsilon(\tau) = \epsilon_i \big( \frac{\tau}{\tau_i} \big)^6 \, ,
 \ea
in which  $\epsilon_i$ is the value of $\epsilon$ at the start of  USR phase. Correspondingly, at the end of USR phase  $\epsilon_e = \epsilon_i \big( \frac{\tau_e}{\tau_i} \big)^6 $. Using  the number of e-fold, $d N= H dt$, the duration of the USR phase is denoted by $\Delta N \equiv N(\tau_e) - N(\tau_i)$ so
$\epsilon_e = e^{-6 \Delta N} \epsilon_i $. 

As shown in \cite{Firouzjahi:2023aum}, a crucial role is played by the sharpness of the transition from the USR phase to the final SR phase. To take this into account, 
following \cite{Cai:2018dkf},  we define the parameter associated to the sharpness of the transition,  $h$, as follows 
\ba
\label{h-def}
h\equiv \frac{6 \sqrt{2 \epsilon_V} }{\dot \phi(t_e)}  = -6 \sqrt{\frac{\epsilon_V}{\epsilon_e}} \, .
\ea
Here, $\epsilon_V$ represents the slow-roll parameter at the final SR phase when the system reaches its attractor regime. 
Since we assume (without lack of generality) that $\phi$ is decreasing during USR phase, then $\dot \phi<0$ so $h<0$.  

As shown in  \cite{Cai:2018dkf} near the transition we can approximate $\eta$ as 
\ba
\eta = -6 - h \theta(\tau -\tau_e) \quad \quad  \tau_e^- < \tau < \tau_e^+ \, .
\ea
In particular, for the derivative of $\eta$, we have 
\ba
\label{eta-jump}
\frac{d \eta}{d \tau} = - h \delta (\tau -\tau_e)  \, ,  \quad \quad  \tau_e^- < \tau < \tau_e^+ \, .
\ea

In the following analysis  we consider two cases of sharp transition: ``natural" sharp transition in which  $\eta$ drops to zero immediately  after transition corresponding to $h=-6$. In this situation   $\epsilon$ after the transition is frozen to its value at the end of USR given by $\epsilon_e$. This limit was studied in  \cite{Kristiano:2022maq, Kristiano:2023scm}. The other case is ``extreme" sharp transition where  $|h| \gg 1$. In this situation, $\epsilon$ after the transition evolves towards the end of inflation (or when the evolution in the final stage has reached its attractor phase) so   $\epsilon_V  = \epsilon_e (\frac{h}{6})^2$. 

As $\epsilon(\tau)$ falls off exponentially during the USR phase, the comoving curvature perturbation $\calR(\tau)$ grows exponentially during the USR phase, $\calR(\tau) \propto a(\tau)^3 \propto \tau^{-3}$.  After the USR period, the curvature perturbation may evolve during the final USR phase until it reaches its final attractor value to be measured at the end of inflation. To read off the final value of $\calR$, we have to track it from the first phase of inflation towards the USR phase and then eventually into the final SR phase. This is achieved by requiring that both $\calR(\tau)$ and $\calR'(\tau)$ to be continuous across the transitions $SR \rightarrow USR \rightarrow SR$. 

Starting with a Bunch-Davies initial condition in the first SR phase, the mode function in the Fourier space  is given by
\ba
\calR^{(1)}_{k} =  \frac{H}{ M_P\sqrt{4 \epsilon_i k^3}} 
( 1+ i k \tau) e^{- i k \tau} \, , \quad \quad (\tau < \tau_i) \, 
\ea 
where $\epsilon_i$ is the value of the slow-roll parameter at the start of inflation 
when the CMB scale mode leaves the horizon. The superscript (1) indicates the first SR phase. During the USR phase, the mode function is given formally by the superposition of the positive and negative frequency modes, 
\ba
\calR^{(2)}_{k} =  \frac{H}{ M_P\sqrt{4 \epsilon_i k^3}}  \big( \frac{\tau_i}{\tau} \big)^3
\Big[ \alpha^{(2)}_k ( 1+ i k \tau) e^{- i k \tau}  + \beta^{(2)}_k ( 1- i k \tau) e^{ i k \tau}  \Big]  \, ,
\ea
with the coefficients $\alpha^{(2)}_k$ and $\beta^{(2)}_k$, after imposing the matching condition at $\tau=\tau_i$, are obtained to be  
\ba
\label{alpha-beta2}
\alpha^{(2)}_k = 1 + \frac{3 i }{ 2 k^3 \tau_i^3} ( 1 + k^2 \tau_i^2) \, , \quad \quad
\beta^{(2)}_k= -\frac{3i }{ 2 k^3 \tau_i^3 } {( 1+ i k \tau_i)^2} e^{- 2 i k \tau_i} \, .
\ea 
Finally, imposing the matching conditions  at $\tau_e$, the mode function in the final SR phase, denoted by the superscript (3), is obtained to be 
\ba
\calR^{(3)}_{k} =  \frac{H}{ M_P\sqrt{4 \epsilon(\tau) k^3}}  
\Big[ \alpha^{(3)}_k ( 1+ i k \tau) e^{- i k \tau}  + \beta^{(3)}_k ( 1- i k \tau) e^{ i k \tau}  \Big] \, ,
\ea
with the coefficients $\alpha^{(3)}_k$ and $\beta^{(3)}_k$  given by,
\ba
\label{alpha-beta3}
\alpha^{(3)}_k = \frac{1}{8 k^6 \tau_i^3 \tau_e^3}  \Big[ 3h
 ( 1 -i k \tau_e)^2 (1+i k \tau_i)^2 e^{2i k (\tau_e- \tau_i)}
+ (-2i k^3 \tau_i^3 + 3 k^2 \tau_i^2 + 3 ) (4 i k^3 \tau_e^3- h k^2 \tau_e^2 - h) \Big]
\nonumber
\ea
and
\ba
\beta^{(3)}_k=   \frac{-1}{8 k^6 \tau_i^3 \tau_e^3}  \Big[ 3 ( 1+ i k \tau_i)^2 ( h+ h k^2 \tau_e^2 + 4 i k^3 \tau_e^3 ) e^{-2 i k \tau_i} -  h ( 1+ i k \tau_e)^2  ( 3  + 3  k^2 \tau_i^2 - 2i k^3 \tau_i^3 ) e^{- 2 i k \tau_e}
 \Big] \nonumber
\ea 

Finally, the power spectrum of curvature perturbations at the end of inflation $\tau=\tau_0 \rightarrow0$ for the mode in the interval $k_i < k< k_e$ which leaves the horizon during the USR phase  is given by
\ba
\label{power-end}
P_\calR(\tau_0, k) = \Big( \frac{h-6}{h} \Big)^2 \frac{H^2}{4 M_P^2 \epsilon_e k^3}
= \Big( \frac{h-6}{h} \Big)^2 P_\calR(\tau_e, k) \, ,   \quad \quad (k_i < k< k_e)\, .
\ea
Curiously, we see that the power spectrum is scaled with a factor $\big( \frac{h-6}{h} \big)^2$ compared to its value at the end of USR phase. In the limit of extreme sharp transition, $h\rightarrow -\infty$, we see that $P_\calR(\tau_0, k) \simeq P_\calR(\tau_e, k)$. This is expected, since in this limit the mode function is frozen immediately after the USR phase and does not experience evolution after the USR phase. 
On the other hand, for the case of natural sharp transition with $h=-6$, we see that $P_\calR(\tau_0, k) \simeq 4P_\calR(\tau_e, k)$ so the power spectrum actually becomes larger towards the end of inflation. This is because the mode function is still evolving after the USR phase until it reaches to its final attractor value. We comment that there are subleading correction of order $O\big( \frac{k^2}{k_e^2}  \big) $
in Eq. (\ref{power-end}) which we have neglected.

On the other hand, the modes which leave the horizon during  the first SR phase are frozen during the intermediate USR phase. Correspondingly, for these modes  (at the tree level) we have 
\ba
\label{power-end2}
P_\calR(\tau_0, k) = \frac{H^2}{4 M_P^2 \epsilon_i k^3}\, ,    \quad \quad
(k < k_i)  \, .
\ea


\section{Cubic and Quartic Hamiltonians  }
\label{eft-action}

\begin{figure}[t]
	\centering
	\includegraphics[ width=0.68\linewidth]{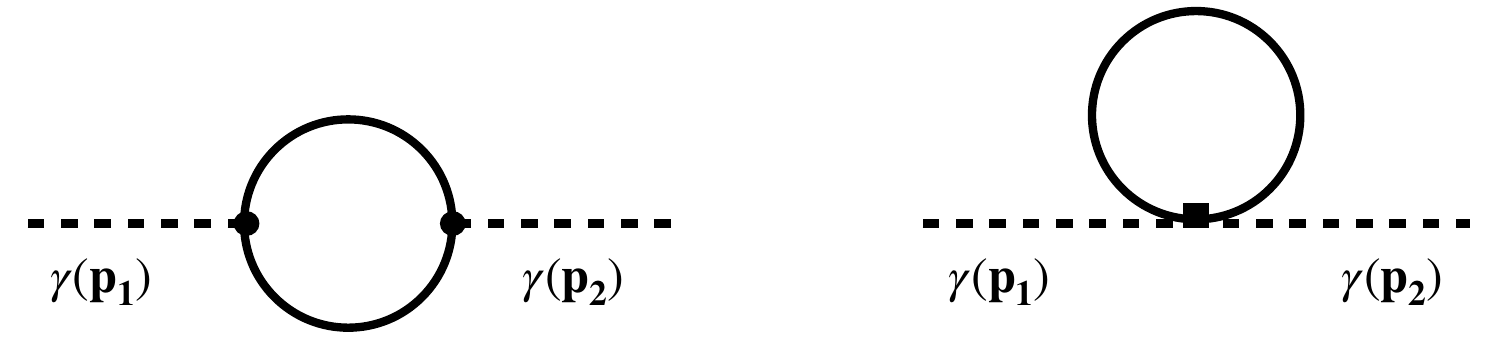}
	\caption{ The Feynman diagrams for the one-loop correction in tensor power spectrum. The dotted line represents the tensor perturbations while the solid line in the loop represents the scalar perturbations. 
The left and right panel represent the contribution  of the cubic  and 
quartic Hamiltonians respectively.  }
\label{Feynman-fig}
\end{figure}

Our goal is to calculate the one-loop corrections in tensor power spectrum induced by the scalar perturbations which experience a growth during the USR phase. 
For this purpose, we need to calculate the cubic and quartic interaction Hamiltonians. Schematically, the cubic Hamiltonian represents an interaction of the type $\gamma \calR^2$ while the quartic Hamiltonian is in the form $\gamma^2 \calR^2$. A schematic view of the corresponding one-loop diagrams associated to these interactions are presented in Fig. \ref{Feynman-fig}. The left panel in Fig. \ref{Feynman-fig} represents the contribution of the cubic Hamiltonian involving a nested in-in integral while the right panel represents the contribution of the quartic Hamiltonian involving a single in-in integral.

We consider the tensor perturbations of the FLRW background as follows, 
\ba
ds^2 = -dt^2 +  g_{i j}d{ x}^i d x^j \, , \quad \quad   g_{ij} \equiv a(t)^2  \hat h_{i j} \, ,
\ea
in which $\hat h_{i j}$ is expanded 
in terms of the tensor perturbations $\gamma_{ij}$ as follows \cite{Maldacena:2002vr}
\ba
\hat h_{i j} = \delta_{ij} + \gamma_{ij} + \frac{1}{2} \gamma_{i \ell} \gamma_{\ell j} 
+ \cdots \, .
\ea
The tensor perturbations are transverse and traceless, $\gamma^{i}_i= \partial_{i} \gamma_{i j}=0$ in which the indices are raised via $\delta^{ij}$. With this construction, there is no  contribution of $\gamma_{ij}$ in $\sqrt{-g}$.

The total action is $S_{\rm total }= S_{\rm matter}+ S_{\rm EH}$ in which $S_{\rm matter}$ is the matter part of the action while $ S_{\rm EH}$ represents the usual Einstein-Hilbert action. To calculate the leading interaction Hamiltonian, we use the effective field theory (EFT) of inflation \cite{Cheung:2007st, Cheung:2007sv}. In a near dS spacetime with a background inflaton field $\phi(t)$, the four-dimensional diffeomorphism invariance  is spontaneously broken to a three-dimensional spatial diffeomorphism invariance. Starting with the unitary (or comoving) gauge
where the perturbations of inflaton are turned off,  
one  is allowed to write down all terms in the action which are consistent with the remaining three-dimensional diffeomorphsim invariance. Upon doing so,  the background inflation dynamics is controlled via the  known  Hubble expansion rate $H(t)$ and its derivative $\dot H(t)$. After writing the full action consistent with the three dimensional diffeomorphsim invariance,  one restores the full four-dimensional diffeomorphsim invariance  by introducing a scalar field fluctuations, $\pi(x^{\mu})$, which is the Goldstone boson associated with the breaking of  the time diffeomorphsim invariance. One big advantage of the EFT approach is when one works in  the decoupling limit where the gravitational back-reactions are neglected.
In this limit  one  neglects the slow-roll suppressed interactions in cubic and quartic actions while keeping only the leading terms which can yield  large non-Gaussianities. In our study concerning the USR setup, these are the interactions  which induce large corrections in one-loop integrals. For earlier work employing EFT approach for the bispectrum analysis in a general non-attractor setup (including the USR setup)  see \cite{Akhshik:2015nfa}. The EFT approach was employed in \cite{Firouzjahi:2023aum} to study the one-loop corrections in scalar power spectrum.

Assuming we have a canonical scalar field with a sound speed $c_s=1$, the matter part of the action   consistent with the  FLRW inflationary background is given by \cite{Cheung:2007st}
 \ba
\label{S-matter}
S_{\rm matter} = \int \! d^4 x  \sqrt{- g} \Big[ -M^2_{P} \dot{H}(t+\pi) \Big(\frac{1}{N^2} (1+\dot\pi-N^i\partial_i\pi)^2-
g^{i j}\partial_i\pi\partial_j\pi \Big)  \nonumber\\
- M^2_{ P} \left(3H^2(t+\pi) +\dot{H}(t+\pi)\right) \Big]  \, ,
\ea
in which  $N$ and $N^i$ are the lapse and shift function in the standard ADM formalism.
In the decoupling limit where the gravitational back-reactions are neglected 
we set $N=1$,  $N^i=0$ and $\sqrt{-g}= a^3$.  Our goal is to read off the interaction between $\pi$ and $\gamma_{ij}$. Since $\gamma_{ij}$ does not contribute into  $\sqrt{-g}$, the coupling between 
$\pi$ and $\gamma_{ij}$ to leading order comes via the interaction $g^{i j}\partial_i\pi\partial_j\pi $.
On the other hand, to quadratic order, we have
\ba
g^{ij}= a^{-2} \big( \delta_{ij} - \gamma_{ij} + \frac{1}{2} 
\gamma_{i \ell} \gamma_{\ell j} \big)  \, ,
\ea
where in the right hand side above, we raise and lower the indices via $\delta_{ij}$. 
Correspondingly, the interaction between $\pi$ and $\gamma_{ij}$ to quartic order has the following terms
\ba
\label{coupling}
g^{i j} \partial_i \pi \partial_j \pi \rightarrow -\gamma_{ij} \partial_i \pi \partial_j \pi 
+ \frac{1}{2} \gamma_{i \ell } \gamma_{\ell j} \partial_i \pi \partial_j \pi \, .
\ea
On the other hand, expanding $\dot H (t +\pi)$ to first order in $\pi$ we have
\ba
\label{dotH}
\dot H (t +\pi)  &=&  \dot H + \ddot H \pi + \cdots , \nonumber\\
&\simeq & - \epsilon H^2 -\epsilon \eta H^3 \pi \, .
\ea
It is important to note that in the USR setup $\eta\simeq -6$, so we can not discard the last term above. 

Plugging Eqs. (\ref{dotH}) and (\ref{coupling}) in the action (\ref{S-matter}) the cubic action is obtained to be  \cite{Noumi:2014zqa}
\ba
S_{\gamma \pi^2} = M_P^2 H^2 \int d \tau d^3 x \,  \epsilon a^2 \gamma_{ij} \partial_i \pi \partial_j \pi \, ,
\ea
 while the quartic action is given by, 
 \ba
 S_{\gamma^2 \pi^2} = M_P^2 H^2 \int d \tau d^3 x \,  \epsilon a^2
 \left[ -\frac{1}{2} \gamma_{i \ell }  \gamma_{\ell j} \partial_i \pi \partial_j \pi 
 + \eta \pi \gamma_{i j} \partial_i \pi \partial_j \pi \right] \, .
 \ea 
Correspondingly, the cubic and quartic interaction Hamiltonians are
\ba
\label{H3}
{\bf H_3} = - M_P^2 H^2 \int  d^3 x \,  \epsilon a^2 \gamma_{ij} \partial_i \pi \partial_j \pi \, ,
\ea
and
\ba
\label{H4}
{\bf H_4} =  M_P^2 H^2 \int  d^3 x \,  \epsilon a^2
 \left[ \frac{1}{2} \gamma_{i \ell }  \gamma_{\ell j} \partial_i \pi \partial_j \pi 
 - \eta  \gamma_{i j} \pi \partial_i \pi \partial_j \pi \right] \, .
\ea
As we see, the quartic Hamiltonian has two terms. One can easily check that the second term above, containing  $ \gamma_{i j}  \pi \partial_i \pi \partial_j \pi $,  does not contribute to graviton power spectrum at one-loop level while it contributes to graviton power spectrum at two-loop level. Therefore, in the following analysis where we study the one-loop correction in graviton power spectrum, we neglect the effects of the second term in ${\bf H_4}$. 

From the above interaction Hamiltonians we see that both ${\bf H_3}$ and ${\bf H_4}$ contain spatial derivatives of the scalar perturbations. This is required because the tensor perturbations carry the indices $i, j$ so they should be contracted with the spatial derivatives of the scalar perturbations. Consequently, one expects that the induced loop corrections in tensor power spectrum  to be suppressed  compared to the case of scalar power spectrum. However, the  amplitude of one-loop corrections in tensor spectrum 
has yet to be calculated.  

Finally, note that curvature perturbations $\calR$ is related to $\pi$ via \cite{Firouzjahi:2023aum}
\ba
 \label{pi-R}
 \calR = - H \pi + O (\pi^2) \, , 
 \ea
in which the higher order terms contain the derivatives of $\pi$ or $H$ \cite{Jarnhus:2007ia, Arroja:2008ga}.  However,  we  calculate the two-point correlation functions at the end of inflation 
 $\tau=\tau_0 \rightarrow 0$ where it is assumed that the system is in the slow-roll regime and the perturbations are frozen on superhorizon scales.  In this case, the higher order corrections in Eq. (\ref{pi-R}) are suppressed and we can simply use the linear relation between $\calR$ and $\pi$ in the following in-in integrals \cite{Firouzjahi:2023aum}. 
 
Going  to Fourier space, the tensor perturbations are expended as follows:
\ba
\label{Fourier}
\gamma_{ij}(x) = \int \frac{d^3 \bfk}{(2 \pi)^3} \sum_{s=\pm } \epsilon_{ij}^s(\bfk) \gamma^s_{\bfk} 
e^{i \bfk\cdot  \bfx} \, ,
\ea  
in which $s=\pm $ are two polarizations of the tensor perturbation. The polarization tensor is transverse and traceless, $\epsilon_{ii}= k^i \epsilon_{ij}=0$ and satisfies 
\ba
\label{polar-conditions}
\epsilon_{ij}^{s*}(\bfk)= \epsilon_{ij}^s(-\bfk)  \, , \quad \quad 
\epsilon_{ij}^s(\bfk) \epsilon_{ij}^{s' *}(\bfk)= 2 \delta_{s s'} \, .
\ea
As an example of polarization tensor, taking $\widehat \bfk$ along the third direction, we choose \cite{Weinberg:2008zzc}
\ba
\label{pol.tensor}
\epsilon_{11}(\hat z, \pm 2)= - \epsilon_{22}(\hat z, \pm 2)= \mp i \epsilon_{12}(\hat z, \pm 2)= \mp i \epsilon_{21}(\hat z, \pm 2) =\frac{1}{\sqrt2},  \quad \epsilon_{i 3}= \epsilon_{3i}=0 \, .
\ea
  
To quantize the free tensor perturbation, as usual we expand the Einstein-Hilbert action to quadratic order in $\gamma_{ij}$ obtaining  \cite{Baumann:2022mni}
\ba
S_{\gamma^2} = \frac{M_P^2}{8} \int d \tau d^3 x\, a^2 \big[ (\gamma_{ij}')^2 - 
(\nabla \gamma_{ij})^2  \big] \, .
\ea
Expanding the quantum operators in terms of the corresponding creation and annihilation operators as,
\ba
\gamma^s_\bfk= b^s_\bfk \gamma_k(\tau) + {b^{s \dagger }_{-\bfk} } \gamma_k(\tau)^* ,
\ea
with the usual commutation relation $[ b^{s_1}_\bfk, b^{s_2}_{\bfk'}] = \delta^{s_1 s_2} \delta^3(\bfk -\bfk')  $, 
the  mode function is given by 
\ba
\gamma_k(\tau) =\frac{H\sqrt2}{M_P k^{\frac{3}{2}}}(1+ i k \tau)e^{-i k \tau}.
\ea
Correspondingly, the  two-point correlation is given by 
\ba
\label{power-gamma}
\langle \gamma^s_\bfk \gamma^{s'}_{{\bfk}'}  \rangle = \frac{ \delta^{s s'}}{2} P_\gamma(k) = \frac{2 H^2}{k^3 M_P^2}  \delta^{s s'} \, ,
\ea  
with the dimensionless tensor power spectrum given by 
\ba
\label{P-gamma}
\calP_\gamma =  \frac{k^3}{2 \pi^2 } P_\gamma(k) = \frac{2 H^2}{\pi^2 M_P^2} \, .
\ea  
  
 To calculate the loop corrections, we employ the standard in-in formalism \cite{Weinberg:2005vy} in which the expectation value of the operator $\widehat {O}$ at the end of inflation $\tau_0$ is given by the Dyson series,
 \ba
 \label{Dyson}
\big  \langle \widehat O(\tau_0)  \big\rangle = \Big \langle \Big[ \bar {\mathrm{T}} \exp \Big( i \int_{-\infty}^{\tau_0} d \tau' H_{in} (\tau') \Big) \Big] \,  \widehat O(\tau_0)  \, \Big[ \mathrm{T} \exp \Big( -i \int_{-\infty}^{\tau_0} d \tau' H_{in} (\tau') \Big) \Big] 
 \Big \rangle \, ,
 \ea
 in which $\mathrm{T}$ and $\bar {\mathrm{T}}$ represent the time ordering and anti-time ordering respectively while $H_{in}(t)$ collectively represents the interaction Hamiltonian. In our case at hand   $H_{in}(\tau) = {\bf H}_3 + {\bf H}_4$.

\section{Tensor-Scalar-Scalar Consistency Condition}
 
 While our main goal is to calculate the one-loop corrections in tensor power spectrum, but as a prelude here we study the bispectrum of $\big \langle \gamma^\lambda_{\bfk_1} \calR_{\bfk_2} \calR_{\bfk_3} \big \rangle $ in the squeezed limit 
 $k_1 \ll k_2 \simeq k_3$. This is mainly to check that our EFT approach with the interaction Hamiltonians given above are trusted for the one-loop corrections in tensor power spectrum. While this analysis in interesting and new (in the current $SR\rightarrow USR\rightarrow SR$ setup), but the reader who is only interested in loop corrections can skip directly to next section.
 
To calculate   $\big \langle \gamma^\lambda_{\bfk_1} \calR_{\bfk_2} \calR_{\bfk_3} \big \rangle $ in the squeezed limit we assume that the tensor perturbation has left the horizon during the first SR phase while the scalar perturbations have left the horizon during the intermediate USR phase. As such, the hierarchy $k_1 \rightarrow 0$ and
$k_2 \simeq k_3$ is assumed.  On the physical ground, as the tensor mode is frozen on superhorizon scale, we expect that a consistency condition similar to that of Maldacena \cite{Maldacena:2002vr} for tensor-scalar-scalar to hold. 
 
To calculate  $\big \langle \gamma^\lambda_{\bfk_1} \calR_{\bfk_2} \calR_{\bfk_3} \big \rangle $ at the tree level, we only need the cubic interaction Hamiltonian ${\bf H_3}$. Plugging ${\bf H_3}$ from Eq. (\ref{H3}) in the in-in integral (\ref{Dyson}), we have
\ba
\label{bispectrum1}
\big \langle \gamma^\lambda_{\bfk_1}(\tau_0) \calR_{\bfk_2}(\tau_0) \calR_{\bfk_3}(\tau_0) \big \rangle 
= - 2 \mathrm {Im} \int_{-\infty}^{\tau_0} d \tau \big \langle {\bf H_3(\tau) }  \gamma^\lambda_{\bfk_1} (\tau_0) \calR_{\bfk_2} (\tau_0)  \calR_{\bfk_3} (\tau_0) 
\big \rangle  \, .
\ea
Using the linear relation $\calR = - H \pi$, and noting that $\bfk_2 \simeq -\bfk_3$, we obtain
\ba
\label{bispectrum2}
\big \langle \gamma^\lambda_{\bfk_1}(\tau_0) \calR_{\bfk_2}(\tau_0) \calR_{\bfk_3}(\tau_0) \big \rangle'  = - 4 M_P^2 \epsilon_{i j}^\lambda(\bfk_1) \widehat \bfk_{2 i}\widehat \bfk_{2 j} \,  {\cal I} \, ,
\ea 
in which here and below a prime over $\langle... \rangle $ means we have pulled out the overall factor $(2 \pi)^3 \delta^3 (\bfk_1 + \bfk_2 + \bfk_3)$. The factor ${\cal I}$ is calculated via the in-in integral as follows,
 \ba
 \label{I-def}
 {\cal I} \equiv  k_2^2 \int_{-\infty}^{\tau_0} d \tau \epsilon(\tau) a^2 \mathrm {Im}\Big[ \gamma_{k_1}^{*}(\tau_0) \calR^*_{k_2}(\tau_0)^2    \gamma_{k_1}(\tau) \calR_{k_2}(\tau)^2   \Big] \, .
 \ea
As the scalar modes leave the horizon during the USR period, there are two contributions in the above integral, from the USR period $\tau_i < \tau < \tau_e$ and after the USR period, $\tau_e < \tau< \tau_0$. Performing the integral over the USR period and neglecting the contribution of a rapidly oscillating term in the form of 
$\cos ( 2 k_2 \tau_i)$,  
we obtain
\ba
{\cal I} (\tau_i < \tau < \tau_e) = -\frac{3}{4} \big( \frac{h-6}{h} \big)^2 \frac{H^4}{k_1^3 k_2^3 M_P^4 \epsilon_e} + {\cal O} \big( \frac{k_2^2}{k_e^2} \big)  \, .
\ea 
On the other hand, calculating ${\cal I}$ for the period $\tau_e < \tau< \tau_0$ we obtain
\ba
{\cal I} (\tau_e < \tau < \tau_0) = -\frac{(6-h) (h-10)}{10 h^2 } 
 \frac{H^4}{k_1^3 k_2^3 M_P^4 \epsilon_e} \times \big( \frac{k_2^2}{k_e^2} \big) \, .
\ea 
For the modes which $k_2 \ll k_e$, we may neglect the contribution ${\cal I} (\tau_e < \tau < \tau_0)$ and to leading order 
\ba
\label{bispectrum3}
\big \langle \gamma^\lambda_{\bfk_1}(\tau_0) \calR_{\bfk_2}(\tau_0) \calR_{\bfk_3}(\tau_0) \big \rangle'  = \frac{3}{4} \epsilon_{i j}^\lambda(\bfk_1) \widehat \bfk_{2 i}\widehat \bfk_{2 j}  P_\calR( k_2, \tau_0)  P_\gamma( k_1, \tau_0)  \, ,
\ea
 in which $P_\calR( k_2, \tau_0) $ and $ P_\gamma( k_1, \tau_0) $ are the scalar and tensor power spectrum as given in Eqs. (\ref{power-end}) and (\ref{power-gamma}). 
 
The above result is obtained employing a direct in-in calculation. However, as the tensor mode is frozen on superhorizon scales and is not affected by the USR phase, we expect a consistency condition similar to \cite{Maldacena:2002vr} to hold. Below we demonstrate that this is indeed the case.
 
As $k_1 \rightarrow 0$, one can assume that the long tensor mode only modifies the background for the short scalar modes \cite{Maldacena:2002vr} in the form a quadrupolar anisotropy by changing $k_2^2 \rightarrow k_2^2 - \gamma_{ij} k_2^i k_2^j$.  Following the logic of \cite{Maldacena:2002vr} we can write
\ba
\label{consistency1}
\big \langle \gamma^\lambda_{\bfk_1} \calR_{\bfk_2}\calR_{\bfk_3} \big \rangle' \simeq  -  \big \langle \gamma^\lambda_{\bfk_1}  \gamma^\lambda_{\bfk_1} \big \rangle \,  \epsilon_{i j}^\lambda(\bfk_1) \widehat \bfk_{2 i}\widehat \bfk_{2 j}  \frac{\partial}{\partial k_2^2}  \langle   \calR_{\bfk_2} \calR_{\bfk_3}    \rangle \, .
\ea
 Using the specific form of the scalar  power spectrum given in Eq. (\ref{power-end})
 we have
 \ba
 \frac{\partial}{\partial k_2^2} P_\calR(k_2) = -\frac{3}{2 k_2^2}P_\calR(k_2) \, ,
 \ea
 and consequently, plugging this in Eq. (\ref{consistency1}), we obtain
 \ba
\label{consistency2}
\big \langle \gamma^\lambda_{\bfk_1}(\tau_0) \calR_{\bfk_2}(\tau_0)\calR_{\bfk_3}(\tau_0) \big \rangle' 
=\frac{3}{4} \epsilon_{i j}^\lambda(\bfk_1) \widehat \bfk_{2 i}\widehat \bfk_{2 j}  P_\calR( k_2, \tau_0)  P_\gamma( k_1, \tau_0)  \, ,
\ea
in exact agreement with Eq. (\ref{bispectrum3}).

As explained above, one expects that the above consistency condition to hold. This is because the tensor perturbation has left the horizon during early SR phase which  
is frozen afterwards and is largely unaffected by the USR phase. Consequently, it can only modify the background for the short scalar modes, which leave the horizon much later in USR phase, in a form of quadrupolar anisotropy.  
 
The above analysis confirms the applicability of our EFT approach. In addition, as the consistency condition is unaffected, the above results imply that the loop corrections from the short  scalar perturbations to be minimal on long tensor perturbations which have left the horizon much earlier. We study this issue more directly in next section.    

\section{Loop Corrections in Tensor Power Spectrum} 
 \label{loop}
 
 Now we study the one-loop corrections in long CMB scale gravitational power spectrum $\langle \gamma^{s_1}(\bfp_1)  \gamma^{s_2}(\bfp_2) \rangle  $ induced from the short scalar modes which leave the horizon during the intermediate USR phase. In our convention  the CMB scale tensor modes  have momentum $\bfp_1$ and $\bfp_2$ while that of short scalar perturbations running in the loop is $\bfq$. 
 
For a consistent one-loop corrections, we have to calculate the contributions of both Feynman diagrams shown in Fig. \ref{Feynman-fig}.  We start with the right panel 
which is easier, containing a four vertex involving one in-in integral over the quartic Hamiltonian ${\bf H_4}$.


\subsection{Loop Corrections from Quartic Hamiltonian} 
\label{loop-quartic}
 
With the quartic Hamiltonian given in Eq. (\ref{H4}) the one-loop correction from the 
right panel of Fig. \ref{Feynman-fig} is given by 
\ba
\big \langle \gamma^{s_1}_{\bfp_{1}}(\tau_0) \gamma^{s_2}_{\bfp_{2}}(\tau_0) \big \rangle_{{\bf H_4}} = 
 - 2 \mathrm {Im} \int_{-\infty}^{\tau_0} d \tau \big \langle\,  {\bf H_4(\tau) }  \, \gamma_{s_1}({\bfp_1} , \tau_0)  \gamma_{s_2}({\bfp_1} , \tau_0)  \, 
\big \rangle  \, ,
\ea
yielding
\ba
 \big \langle \gamma^{s_1}_{\bfp_{1}}(\tau_0) \gamma^{s_2}_{\bfp_{2}}(\tau_0) \big \rangle'_{{\bf H_4}}
 = - 2 M_P^2\,  \mathrm{Im} \Big[  \epsilon^{s_1}_{i \ell} (-\bfp_1)  
 \epsilon^{s_2}_{ \ell j} (\bfp_1)    \int \frac{d^3 \bfq}{(2 \pi)^3} \,  q_i q_j  \,   \,  {\cal I}_4 (q)\Big] \, ,
\ea 
in which the factor ${\cal I}_4(q)$ associated to the quartic Hamiltonian in-in integral 
will be given shortly below. 

Using the isotropy of the background, the integral $\int { d^3 \bfq}  \,  q_i q_j  {\cal I}_4 (q)$ is non-zero only $i=j$ so one can replace this momentum integral by $\frac{1}{3}\delta_{ij}\int { d^3 \bfq}  \,  q^2 {\cal I}_4 (q)$. Now  using the properties of the polarization tensor given in Eq. (\ref{polar-conditions}) we obtain
\ba
\epsilon^{s_1}_{i \ell} (-\bfp_1)  
 \epsilon^{s_2}_{ \ell j} (\bfp_1) \int \frac{ d^3 \bfq}{(2 \pi)^3}  \,  q_i q_j  {\cal I}_4 (q)   = \frac{2}{3}  \delta_{s_1 s_2}  \int  \frac{ d^3 \bfq}{(2 \pi)^3}  \,  q^2  {\cal I}_4 (q)  \, .
\ea
Combining all together, we obtain
\ba
\label{power-quartic1}
  \big \langle \gamma^{s_1}_{\bfp_{1}}(\tau_0) \gamma^{s_2}_{\bfp_{2}}(\tau_0) \big \rangle'_{{\bf H_4}}
 = - \frac{4 \delta^{s_1 s_2}}{3} M_P^2\,  \int\frac{ d^3 \bfq}{(2 \pi)^3}  \,  q^2\, 
 \mathrm{Im} \,  {\cal I}_4 (q) \, ,
 \ea
 in which the factor ${\cal I}_4(q)$ is given by 
 \ba
 \label{I4-def}
{\cal I}_4 (q) \equiv   \int_{-\infty}^{\tau_0} d \tau \epsilon(\tau) a^2   \Big[ \gamma^*(p_1, \tau_0)^2 \gamma(p_1, \tau)^2
\Big]  \big| \calR(q, \tau) \big|^2 \, .
\ea 
In performing the time  integral  above we should only consider the contribution of the superhorizon modes, so the actual time interval in Eq. (\ref{I4-def}) should be $-\frac{1}{q} < \tau < \tau_0$. This guarantees that we do not count the contributions of the modes which are subhorizon (i.e. not yet classical) in the time integral in Eq. (\ref{I4-def}). On the other hand, the modes which are subhorizon during the  USR phase are quantum mechanical in nature so their contributions should be collected via a UV renormalization scheme. While renormalization is an important issue to read off the final physical quantity but here we are mainly interested in the effects of superhorizon modes to obtain a rough estimate for the magnitude of the loop corrections. In addition,  as $q\tau \rightarrow 0 $, the integral  in Eq. (\ref{I4-def}) receives its contribution from its lower end. In particular, the contribution from the period after the USR phase  $\tau_e < \tau < \tau_0$ is subleading.
 
In the limit that $p \rightarrow 0$, we have  
\ba
\label{Im1}
\mathrm{Im} \Big[ \gamma^*(p_1, \tau_0)^2 \gamma(p_1, \tau)^2
\Big] \simeq -\frac{8}{3} \frac{H^4 \tau^3}{M_P^4 p^3} \, .
\ea
Furthermore, on the superhorizon in which $q \tau \rightarrow 0$, we have 
$\epsilon(\tau) \big| \calR(q, \tau) \big|^2 \simeq \frac{H^2}{4 q^3 M_P^2}$, yielding
\ba
\mathrm{Im}\,   {\cal I}_4 (q) \simeq -\frac{2 H^4 }{3 M_P^6 q^3 p^3} \int_{-\frac{1}{q}}^{\tau_0} d \tau  \tau  \simeq \frac{ H^4 }{3 M_P^6 q^5 p^3} \, .
\ea
Plugging the above result in Eq. (\ref{power-quartic1}) and integrating over the USR modes $q_i < q< q_e$,  we obtain 
\ba
\label{loop1}
  \big \langle \gamma^{s_1}_{\bfp_{1}}(\tau_0) \gamma^{s_2}_{\bfp_{2}}(\tau_0) \big \rangle'_{{\bf H_4}}
 \simeq  - \frac{4 \delta^{s_1 s_2}}{9} \frac{ H^4 }{ M_P^4  p^3} \frac{\Delta N}{2 \pi^2} \, ,
\ea
in which $\Delta N = \ln \big( \frac{\tau_i}{\tau_e} \big) $ is the duration of the USR phase.

It is convenient to express the  loop correction in terms of the dimensionless 
power spectrum $\calP_\gamma$ defined in Eq. (\ref{P-gamma}). Using the result from Eq. (\ref{loop1}), for the one-loop correction in tensor power spectrum 
from the quartic Hamiltonian 
${\bf H_4}$ we obtain 
\ba
\label{power-quartic}
\calP_\gamma^{(\mathrm{loop})}\big|_{{\bf H_4}} \simeq -\frac{\Delta N}{36} \calP_\gamma^2 \, .
\ea 

Since we calculate the loop corrections induced from the scalar perturbations on tensor power spectrum, then one expects the  loop correction to scale like
$\calP_\gamma \calP_\calR$. However, from Eq. (\ref{power-quartic}) we see that the loop correction actually scales like $\calP_\gamma^2$. The reason is that the interaction vertices in ${\bf H_3}$ and ${\bf H_4} $
contain the factor $\epsilon$ so the combination $\epsilon \calR^2$ appears inside the in-in integral as in Eq. (\ref{I4-def}). Since $\epsilon \calR^2 \sim \calP_\gamma$, then the  final result for loop correction is given as $\calP_\gamma^2$ instead of $\calP_\gamma \calP_\calR$.

\subsection{Loop Corrections from Cubic Hamiltonian} 
 \label{loop-cubic} 
 
 Now we calculate the loop corrections from the cubic Hamiltonian corresponding to the left panel of Fig. \ref{Feynman-fig}. It involves a nested integral containing the product of two three-vertices. More schematically, expanding the Dyson series to second order in $\Ha$ we have 
\ba
\langle \gamma^{s_1}_{\bfp_{1}}(\tau_0) \gamma^{s_2}_{\bfp_{2}}(\tau_0) \rangle_{\Ha} =    \langle \gamma^{s_1}_{\bfp_{1}}(\tau_0) \gamma^{s_2}_{\bfp_{2}}(\tau_0) \rangle_{(2,0)} +   \langle\gamma^{s_1}_{\bfp_{1}}(\tau_0) \gamma^{s_2}_{\bfp_{2}}(\tau_0) \rangle_{(1,1)} +    \langle \gamma^{s_1}_{\bfp_{1}}(\tau_0) \gamma^{s_2}_{\bfp_{2}}(\tau_0) \rangle_{(0, 2)}
  \ea
  in which 
\ba
  \label{20-int}
\big \langle \gamma^{s_1}_{\bfp_{1}}(\tau_0) \gamma^{s_2}_{\bfp_{2}}(\tau_0) \big \rangle_{(2,0)} &=&
- \int_{-\infty}^{\tau_0} d \tau_1 \int_{-\infty}^{\tau_1} d \tau_2
\big \langle \Ha (\tau_2)  \Ha (\tau_1) \gamma^{s_1}_{\bfp_{1}}(\tau_0) \gamma^{s_2}_{\bfp_{2}}(\tau_0)
 \big \rangle   \nonumber\\
&=& \big \langle  \gamma^{s_1}_{\bfp_{1}}(\tau_0) \gamma^{s_2}_{\bfp_{2}}(\tau_0) 
\big \rangle^\dagger_{(0,2)}\, ,
\ea
and 
 \ba
  \label{11-int}
\langle \gamma^{s_1}_{\bfp_{1}}(\tau_0) \gamma^{s_2}_{\bfp_{2}}(\tau_0) \rangle_{(1,1)} =
 \int_{-\infty}^{\tau_0} d \tau_1 \int_{-\infty}^{\tau_0} d \tau_2 \, 
\big \langle \Ha (\tau_1)  \gamma^{s_1}_{\bfp_{1}}(\tau_0) \gamma^{s_2}_{\bfp_{2}}(\tau_0)
\Ha (\tau_2) \big \rangle   \, .
\ea

We leave the details of the in-in analysis into Appendix. After a long calculation, one obtains 
\ba
\label{loop-cubic1}
\langle \gamma^{s_1}_{\bfp_{1}}(\tau_0) \gamma^{s_2}_{\bfp_{2}}(\tau_0) \rangle'_{\Ha} = - 8 M_P^4 \delta^{s_1 s_2}  \int  \frac{ d^3 \bfq}{(2 \pi)^3} 
\big |  \epsilon^{s_1}_{i j} ( \bfp)  q_i  q_j \big|^2 \int_{-\infty}^{\tau_0} d \tau_1 
\int_{-\infty}^{\tau_1} d \tau_2 \mathrm{Im}\,  \big[ X^*(\tau_2) \delta (\tau_1)
\big] \, ,
\ea 
in which
\ba
\label{X-def}
X(\tau) \equiv \epsilon a^2 \gamma (p, \tau) \gamma^{ *} (p, \tau_0)\calR(q, \tau)^2   \, ,
\ea 
and
\ba
\label{delta-def}
\delta(\tau) \equiv 2 \epsilon a^2  \calR(q, \tau)^2 \mathrm{Im} \big [  \gamma (p, \tau) \gamma^{ *} (p, \tau_0)  \big] \, .
\ea
 Using the orthogonality properties of the polarization tensor one can show that
 \ba
 \int d \Omega \big |  \epsilon^{s_1}_{i j} ( \bfp)  \hat{q_i}  \hat{q_j} \big|^2 =\frac{16 \pi}{15} \, .
 \ea
in which $d \Omega$ represents  the angular parts of $d^3 \bfq$. 
 
Combining all contributions, we obtain (see Appendix for further details)
\ba
\label{power-cubic}
\calP_\gamma^{(\mathrm{loop})} \big|_{{\bf H_3}}  \simeq - \frac{19}{720}  \Delta N  \calP_\gamma^2 \, .
\ea 
In obtaining the above result, we have neglected the terms without the factor $\Delta N$
which appears as subleading correction for large enough $\Delta N$.
 As in quartic case the loop correction scales like $\calP_\gamma^2$ instead of $ \calP_\calR \calP_\gamma$.


Now combining the results from the cubic and quartic interactions, Eqs. (\ref{power-cubic}) and (\ref{power-quartic}), 
 the total one-loop correction is obtained to be\footnote{There is a mistake in the original version of this analysis containing a factor $e^{\Delta N}$ in $\calP_\gamma^{(\mathrm{loop})}$. This is due to the mistake in the positions of $\tau_1$ and $\tau_2$ in Eq. (\ref{20-int})
in the original version of this analysis.}
\ba
\label{power-total}
\calP_\gamma^{(\mathrm{loop})}    \simeq -\frac{13 \Delta N}{240}  \calP_\gamma^2 \, ,
\ea 
with subleading corrections without factor of $\Delta N$, i.e. the subleading term of the form $c \, \calP_\gamma^2$ with the numerical factor $c$ independent of $\Delta N$.   

From the above result we see that the loop corrections in tensor power spectrum induced from the USR modes are quite insensitive to the sharpness of the transition from the USR phase to the SR phase. Indeed, we do not see any explicit dependence to the sharpness parameter $h$ in Eq. (\ref{power-total}). This is unlike the loop corrections induced on long scalar perturbations  in which the loop corrections increase linearly with $h$ \cite{Firouzjahi:2023aum} for $|h |\gg 1$  in which  $\calP_\calR^{(\mathrm{loop})}   \sim h \calP_\calR^{\mathrm{CMB}} \,   \calP_\calR^{\mathrm{short}}  \sim h \big( \calP_\calR^{\mathrm{CMB}} \big)^2\,   e^{6 \Delta N}$.  In addition, we see that the induced loop corrections in GWs 
are quite small. More specifically  we obtain $\frac{\calP_\gamma^{(\mathrm{loop})}}{\calP_\gamma} \sim 10^{-2} \, \Delta N \calP_\gamma$. Assuming $\calP_\gamma \lesssim 10^{-10}$ from the CMB observations, we see that the fractional loop corrections in tensor perturbation is smaller than $10^{-14}$.

The conclusion is that the long CMB scale gravitational waves are practically unaffected by  the short scalar perturbations which leave the horizon during the
USR phase. This conclusion is largely  independent of the mechanism of the transition from the USR phase to the final SR phase.

\section{Summary and Discussions} 
  
In this work we have studied the one-loop correction in power spectrum of long gravitational waves from small scale modes which leave the horizon during the intermediate USR phase. This study is motivated by similar  recent studies  performed for loop corrections in scalar power spectrum.

As one might have guessed, the results are quite different from what were obtained for the case of scalar power spectrum. We have shown that the long tensor power spectrum is largely unaffected by the loop corrections from small USR modes. In particular, the one-loop corrections are quite insensitive to the sharpness of the transition. This might have been expected from the physical  ground that the tensor perturbations only probe the Hubble expansion rate of the corresponding inflationary background and are insensitive to slow-roll parameters. Having said this, it is still a good cross check to verify the validity of this physical expectation 
since a similar intuition, suggesting that the scalar power spectrum should be unaffected by intermediate short modes, proved to fail for the case of a sharp transition \cite{Kristiano:2022maq, Kristiano:2023scm,  Firouzjahi:2023aum}. 
While our analysis was focused to the particular setup of $SR\rightarrow USR \rightarrow SR$, but this conclusion may be general. As long as there is no dramatic changes in the background Hubble expansion rate, then independent of the nature of transitions in slow-roll parameters, the superhorizon tensor modes are unaffected by the short scalar modes which may experience rapid growth. It would be useful to verify this conjecture in its generality. 

In addition we have shown that the  Maldacena consistency condition for the tensor-scalar-scalar bispectrum in the squeezed limit does hold.  The fact that the long tensor mode is frozen on superhorizon scale is the key reason for the validity of this consistency condition. The long tensor perturbations only induce small anisotropies on the background for the short modes yielding to the expected tensor-scalar-scalar consistency condition.

We comment that the loop corrections on tensor power spectrum calculated here should not be confused with the induced gravitational waves from second order scalar perturbations which are actively investigated recently, for a review see \cite{Domenech:2021ztg} and for  works studying  secondary GWs induced in 
models with non-Gaussian feature or USR  setup see \cite{Cai:2018dig, Liu:2020oqe, Talebian:2022cwk, Ragavendra:2021qdu}. While these two questions are related but the induced GWs from large second order    scalar perturbations are mostly concerned with small scale GWs, the modes  near the peak of scalar perturbations,  
which re-enter the horizon during the radiation dominated era. Here, on the other hand, we look at the enhancement of GWs spectrum at the CMB scales. 
  
 \vspace{1cm}
   
 {\bf Acknowledgments:}  I am grateful to  Mohammad Ali Gorji and Antonio Riotto  
for useful  discussions and for comments on the draft.  I would like to thank Amin Nassiri-Rad for checking the in-in analysis in Section 5. 
 This research is  partially supported by the ``Saramadan" Federation of Iran.

\appendix 
\section{In-In Analysis for  Cubic Hamiltonian}
\label{Appendix}

In this Appendix we present the details of the in-in integral for the cubic Hamiltonians
${\bf H_3}$. 

As discussed before, the loop interaction from the cubic Hamiltonian is given by 
\ba
\label{cubic-in-in}
\langle \gamma^{s_1}_{\bfp_{1}}(\tau_0) \gamma^{s_2}_{\bfp_{2}}(\tau_0) \rangle_{\Ha} =    \langle \gamma^{s_1}_{\bfp_{1}}(\tau_0) \gamma^{s_2}_{\bfp_{2}}(\tau_0) \rangle_{(2,0)} +   \langle\gamma^{s_1}_{\bfp_{1}}(\tau_0) \gamma^{s_2}_{\bfp_{2}}(\tau_0) \rangle_{(1,1)} +    \langle \gamma^{s_1}_{\bfp_{1}}(\tau_0) \gamma^{s_2}_{\bfp_{2}}(\tau_0) \rangle_{(0, 2)}
  \ea
with
\ba
  \label{20-int-b}
\big \langle \gamma^{s_1}_{\bfp_{1}}(\tau_0) \gamma^{s_2}_{\bfp_{2}}(\tau_0) \big \rangle_{(2,0)} &=&
- \int_{-\infty}^{\tau_0} d \tau_1 \int_{-\infty}^{\tau_1} d \tau_2
\big \langle \Ha (\tau_2)  \Ha (\tau_1) \gamma^{s_1}_{\bfp_{1}}(\tau_0) \gamma^{s_2}_{\bfp_{2}}(\tau_0)
 \big \rangle   \nonumber\\
&=& \big \langle  \gamma^{s_1}_{\bfp_{1}}(\tau_0) \gamma^{s_2}_{\bfp_{2}}(\tau_0) 
\big \rangle^\dagger_{(0,2)}\, ,
\ea
and 
 \ba
  \label{11-int-b}
\langle \gamma^{s_1}_{\bfp_{1}}(\tau_0) \gamma^{s_2}_{\bfp_{2}}(\tau_0) \rangle_{(1,1)} =
 \int_{-\infty}^{\tau_0} d \tau_1 \int_{-\infty}^{\tau_0} d \tau_2 \, 
\big \langle \Ha (\tau_1)  \gamma^{s_1}_{\bfp_{1}}(\tau_0) \gamma^{s_2}_{\bfp_{2}}(\tau_0)
\Ha (\tau_2) \big \rangle   \, .
\ea

Let us start with $\langle\gamma^{s_1}_{\bfp_{1}}(\tau_0) \gamma^{s_2}_{\bfp_{2}}(\tau_0) \rangle_{(1,1)}$. Using the Hamiltonian (\ref{H3}), performing all contractions and employing the properties of the polarization tensor given in Eq. (\ref{polar-conditions}) one obtains
\ba
\label{11-int}
\langle \gamma^{s_1}_{\bfp_{1}}(\tau_0) \gamma^{s_2}_{\bfp_{2}}(\tau_0) \rangle'_{(1,1)}= 4 M_P^4  \delta^{s_1 s_2}  \int  \frac{ d^3 \bfq}{(2 \pi)^3} 
\big |  \epsilon^{s_1}_{i j} ( \bfp)  q_i  q_j \big|^2\,   \Big|\int_{-\infty}^{\tau_0} d \tau X(\tau)  \Big|^2 \, ,
\ea
in which 
\ba
\label{X-def2}
X(\tau) \equiv \epsilon a^2 \gamma (p, \tau) \gamma^{ *} (p, \tau_0)\calR(q, \tau)^2   \, .
\ea 
Similarly, for $\big \langle \gamma^{s_1}_{\bfp_{1}}(\tau_0) \gamma^{s_2}_{\bfp_{2}}(\tau_0) \big \rangle'_{(2,0)}$ we obtain
\ba
\big \langle \gamma^{s_1}_{\bfp_{1}}(\tau_0) \gamma^{s_2}_{\bfp_{2}}(\tau_0) \big \rangle'_{(2,0)} = -4 M_P^4  \delta^{s_1 s_2}  \int  \frac{ d^3 \bfq}{(2 \pi)^3} 
\big |  \epsilon^{s_1}_{i j} ( \bfp)  q_i  q_j \big|^2\,  \int_{-\infty}^{\tau_0} d \tau_1 Z(\tau_1)   \int_{-\infty}^{\tau_1} d \tau_2 X(\tau_2) \, ,
\ea
in which
\ba
\label{Z-def}
Z(\tau) \equiv \epsilon a^2 \gamma (p, \tau) \gamma^{ *} (p, \tau_0)\calR^*(q, \tau)^2 \, .
\ea
Noting that $\big \langle \gamma^{s_1}_{\bfp_{1}}(\tau_0) \gamma^{s_2}_{\bfp_{2}}(\tau_0) \big \rangle_{(2,0)} =  \big \langle  \gamma^{s_1}_{\bfp_{1}}(\tau_0) \gamma^{s_2}_{\bfp_{2}}(\tau_0) 
\big \rangle^\dagger_{(0,2)}$, we obtain 
\ba
\label{nested1}
\big \langle \gamma^{s_1}_{\bfp_{1}}(\tau_0) \gamma^{s_2}_{\bfp_{2}}(\tau_0) \big \rangle'_{(2,0)} +  \big \langle  \gamma^{s_1}_{\bfp_{1}}(\tau_0) \gamma^{s_2}_{\bfp_{2}}(\tau_0)\big \rangle'_{(0,2)} &= &
-4 M_P^4  \delta^{s_1 s_2}  \int  \frac{ d^3 \bfq}{(2 \pi)^3} 
\big |  \epsilon^{s_1}_{i j} ( \bfp)  q_i  q_j \big|^2\, \\
&\times &
 \int_{-\infty}^{\tau_0} d \tau_1 
 \int_{-\infty}^{\tau_1} d \tau_2 \big[  X(\tau_2)   Z(\tau_1) + X^*(\tau_2)   Z^*(\tau_1)
 \big] \, . \nonumber
\ea
To proceed further, let us define
\ba
Z(\tau) \equiv X^*(\tau) + i \delta(\tau)^* \, ,
\ea
in which the new variable $\delta$, from Eqs. (\ref{Z-def}) and (\ref{X-def2}), is obtained to be
\ba
\label{delta-def2}
\label{delta-def}
\delta(\tau) =  2 \epsilon a^2  \calR(q, \tau)^2 \mathrm{Im} \big [  \gamma (p, \tau) \gamma^{ *} (p, \tau_0)  \big] \, .
\ea
With the above relation  between $X(\tau)$ and $Z(\tau)$, one can show that the nested time integrals in Eq. (\ref{nested1}) is rearranged in the following form
\ba
\label{nested2}
\int_{-\infty}^{\tau_0} d \tau_1 
 \int_{-\infty}^{\tau_1} d \tau_2 \big[  X(\tau_1)   Z(\tau_2) + X^*(\tau_1)   Z^*(\tau_2)
 \big] &=& \int_{-\infty}^{\tau_0} d \tau  \big|X(\tau)\big|^2 \\
 &-2&  \int_{-\infty}^{\tau_0} d \tau_1  \int_{-\infty}^{\tau_1} d \tau_2 \mathrm {Im} \big[ X(\tau_2) \delta^*(\tau_1) \nonumber
 \big] \, .
\ea
We see that the first integral in Eq. (\ref{nested2}) cancels the contribution of $\langle \gamma^{s_1}_{\bfp_{1}}(\tau_0) \gamma^{s_2}_{\bfp_{2}}(\tau_0) \rangle'_{(1,1)}$ in Eq. (\ref{11-int}) so at the end we are left with
\ba
\label{loop-cubic2}
\langle \gamma^{s_1}_{\bfp_{1}}(\tau_0) \gamma^{s_2}_{\bfp_{2}}(\tau_0) \rangle'_{\Ha} = - 8 M_P^4 \delta^{s_1 s_2}  \int  \frac{ d^3 \bfq}{(2 \pi)^3} 
\big |  \epsilon^{s_1}_{i j} ( \bfp)  q_i  q_j \big|^2 \int_{-\infty}^{\tau_0} d \tau_1 
\int_{-\infty}^{\tau_1} d \tau_2 \mathrm{Im}\,  \big[ X^*(\tau_2) \delta (\tau_1)
\big] \, .
\ea 

To go further, we need to calculate the contribution of the polarization tensor in the above integral.  With the specific form of the polarization tensor given in Eq. (\ref{pol.tensor}), one can show that 
\ba
 \epsilon^{\pm}_{i j} ( \bfp)  \widehat {q_i } \widehat{ q_j} =\frac{1}{\sqrt2} \sin^2(\theta) e^{ \pm 2 i \phi} \, ,
\ea
in which the orientation of the unit vector $\widehat q$ in a coordinate where $\widehat \bfp$ is along the third axis is specified by the angles $(\phi, \theta)$ in which $\widehat q= (\sin \theta \cos \phi, \sin \theta \sin\phi, \cos \theta)$. 
Consequently, one can easily check that
\ba
\int d \Omega  \big( \epsilon^{s_1}_{i j} ( \bfp)  \widehat {q_i } \widehat{ q_j} \big)
\big( \epsilon^{s_2 *}_{m n} ( \bfp)  \widehat{q_m} \widehat{q_n} \big) =
 \frac{16 \pi}{15} \delta^{s_1 s_2} \, .
\ea
Plugging the above result in Eq. (\ref{loop-cubic2}) we obtain 
\ba
\label{loop-cubic3}
\langle \gamma^{s_1}_{\bfp_{1}}(\tau_0) \gamma^{s_2}_{\bfp_{2}}(\tau_0) \rangle'_{\Ha} = - \frac{16}{15 \pi^2} M_P^4 \delta^{s_1 s_2}  \int  dq 
q^6 \int_{-\infty}^{\tau_0} d \tau_1 
\int_{-\infty}^{\tau_1} d \tau_2 \mathrm{Im}\,  \big[ X^*(\tau_2) \delta (\tau_1)
\big] \, .
\ea 
In performing the above nested integral, it is useful to note that
\ba
\label{app1}
\mathrm{Im} \big [  \gamma (p, \tau) \gamma^{ *} (p, \tau_0)  \big] 
=-\frac{2 H^2}{3 M_P^2} \tau^3 \, ,
\ea
and
\ba
\label{app2}
\gamma (p, \tau) \gamma^{ *} (p, \tau_0) = \frac{2 H^2}{ M_P^2 p^3}
+ {\cal O}(p^{-1} ) \, .
\ea

There is an important  comment in order. We emphasis that we integrate over the modes which become superhorizon during the USR phase, so the time integrals in Eq. (\ref{loop-cubic3}) are actually restricted to
$-\frac{1}{q} < \tau_2 < \tau_1 < \tau_e$. This it to make sure that we only count the modes which become classical during the USR phase. The modes which are subhorizon during the  USR phase are not classical and their effects may be collected under a UV renormalization scheme which is not our question of interest here. With the same logic, for the integral over the momentum $q$ we integrate over  the modes  $q_i < q < q_e$ which become superhorizon during the USR phase. 

Using the relations (\ref{app1}) and (\ref{app2}) for $\delta (\tau_1)$ and $X(\tau_2)$ in the nested integral (\ref{loop-cubic3}) we obtain Eq. (\ref{power-cubic}) in the main text. We comment that the main contribution in the time integral 
in Eq. (\ref{loop-cubic3}) comes for the USR period, $\tau_i < \tau_{1,2} < \tau_e$
while the contribution from the final SR phase, $\tau_e < \tau_{1,2} < \tau_0$,  is subleading.  To perform the analysis of the nested integral in Eq. (\ref{loop-cubic3}) we use the Maple computational software.


{}


\begin{thebibliography}{}




\bibitem{Kristiano:2022maq}
J.~Kristiano and J.~Yokoyama,
[arXiv:2211.03395 [hep-th]].

\bibitem{Kristiano:2023scm}
J.~Kristiano and J.~Yokoyama,
[arXiv:2303.00341 [hep-th]].

\bibitem{Riotto:2023gpm}
A.~Riotto,
[arXiv:2303.01727 [astro-ph.CO]].

\bibitem{Riotto:2023hoz}
A.~Riotto,
[arXiv:2301.00599 [astro-ph.CO]].

\bibitem{Choudhury:2023vuj}
S.~Choudhury, M.~R.~Gangopadhyay and M.~Sami,
[arXiv:2301.10000 [astro-ph.CO]].

\bibitem{Choudhury:2023jlt}
S.~Choudhury, S.~Panda and M.~Sami,
[arXiv:2302.05655 [astro-ph.CO]].

\bibitem{Choudhury:2023rks}
S.~Choudhury, S.~Panda and M.~Sami,
[arXiv:2303.06066 [astro-ph.CO]].


\bibitem{Firouzjahi:2023aum}
H.~Firouzjahi,
[arXiv:2303.12025 [astro-ph.CO]].
  
\bibitem{Motohashi:2023syh}
H.~Motohashi and Y.~Tada,
[arXiv:2303.16035 [astro-ph.CO]].

\bibitem{Firouzjahi:2023ahg}
H.~Firouzjahi and A.~Riotto,
[arXiv:2304.07801 [astro-ph.CO]].

\bibitem{Cheng:2021lif}
S.~L.~Cheng, D.~S.~Lee and K.~W.~Ng,
Phys. Lett. B \textbf{827}, 136956 (2022). 

\bibitem{Ivanov:1994pa}
P.~Ivanov, P.~Naselsky and I.~Novikov,
Phys. Rev. D \textbf{50}, 7173-7178 (1994). 

\bibitem{Garcia-Bellido:2017mdw}
J.~Garcia-Bellido and E.~Ruiz Morales,
Phys. Dark Univ. \textbf{18}, 47-54 (2017). 

\bibitem{Biagetti:2018pjj}
M.~Biagetti, G.~Franciolini, A.~Kehagias and A.~Riotto,
JCAP \textbf{07}, 032 (2018). 

\bibitem{Ozsoy:2023ryl}
O.~\"Ozsoy and G.~Tasinato,
[arXiv:2301.03600 [astro-ph.CO]].

\bibitem{Byrnes:2021jka}
C.~T.~Byrnes and P.~S.~Cole,
[arXiv:2112.05716 [astro-ph.CO]].


\bibitem{Kinney:2005vj} 
  W.~H.~Kinney,
  Phys.\ Rev.\ D {\bf 72}, 023515 (2005)
  [gr-qc/0503017].
  
\bibitem{Morse:2018kda}
M.~J.~P.~Morse and W.~H.~Kinney,
Phys. Rev. D \textbf{97}, no.12, 123519 (2018). 

\bibitem{Lin:2019fcz}
W.~C.~Lin, M.~J.~P.~Morse and W.~H.~Kinney,
JCAP \textbf{09}, 063 (2019). 

  
\bibitem{Namjoo:2012aa} 
  M.~H.~Namjoo, H.~Firouzjahi and M.~Sasaki,
  Europhys.\ Lett.\  {\bf 101}, 39001 (2013).

\bibitem{Maldacena:2002vr} 
  J.~M.~Maldacena,
  JHEP {\bf 0305}, 013 (2003)
  [astro-ph/0210603].

\bibitem{Creminelli:2004yq}
P.~Creminelli and M.~Zaldarriaga,
JCAP \textbf{10}, 006 (2004). 


\bibitem{Martin:2012pe}
J.~Martin, H.~Motohashi and T.~Suyama,
Phys. Rev. D \textbf{87}, no.2, 023514 (2013). 
  
  
\bibitem{Chen:2013aj} 
  X.~Chen, H.~Firouzjahi, M.~H.~Namjoo and M.~Sasaki,
  Europhys.\ Lett.\  {\bf 102}, 59001 (2013). 
 
\bibitem{Chen:2013eea} 
  X.~Chen, H.~Firouzjahi, E.~Komatsu, M.~H.~Namjoo and M.~Sasaki,
  JCAP {\bf 1312}, 039 (2013). 
  
\bibitem{Akhshik:2015rwa}
M.~Akhshik, H.~Firouzjahi and S.~Jazayeri,
JCAP \textbf{12}, 027 (2015). 

\bibitem{Mooij:2015yka}
S.~Mooij and G.~A.~Palma,
JCAP \textbf{11}, 025 (2015). 

\bibitem{Bravo:2017wyw}
R.~Bravo, S.~Mooij, G.~A.~Palma and B.~Pradenas,
JCAP \textbf{05}, 024 (2018). 

\bibitem{Finelli:2017fml}
B.~Finelli, G.~Goon, E.~Pajer and L.~Santoni,
Phys. Rev. D \textbf{97}, no.6, 063531 (2018). 

\bibitem{Pi:2022ysn}
S.~Pi and M.~Sasaki,
Phys. Rev. Lett. \textbf{131}, no.1, 011002 (2023). 

\bibitem{Cai:2018dkf}
Y.~F.~Cai, X.~Chen, M.~H.~Namjoo, M.~Sasaki, D.~G.~Wang and Z.~Wang,
JCAP \textbf{05}, 012 (2018). 


\bibitem{Ota:2022xni}
A.~Ota, M.~Sasaki and Y.~Wang,
[arXiv:2211.12766 [astro-ph.CO]].

\bibitem{Chen:2022dah}
C.~Chen, A.~Ota, H.~Y.~Zhu and Y.~Zhu,
Phys. Rev. D \textbf{107}, no.8, 083518 (2023). 

\bibitem{Ota:2022hvh}
A.~Ota, M.~Sasaki and Y.~Wang,
[arXiv:2209.02272 [astro-ph.CO]].

\bibitem{Meng:2022ixx}
D.~S.~Meng, C.~Yuan and Q.~g.~Huang,
Phys. Rev. D \textbf{106}, no.6, 063508 (2022). 

\bibitem{Brahma:2022yxu}
S.~Brahma, A.~Berera and J.~Calder\'on-Figueroa,
JHEP \textbf{08}, 225 (2022). 


\bibitem{Weinberg:2008zzc}
S.~Weinberg,
``Cosmology,'' Oxford University Press, 2008. 


\bibitem{Baumann:2022mni}
D.~Baumann,
``Cosmology,'' Cambridge University Press, 2022,

\bibitem{Kodama:1984ziu}
H.~Kodama and M.~Sasaki,
Prog. Theor. Phys. Suppl. \textbf{78}, 1-166 (1984). 


\bibitem{Mukhanov:1990me}
V.~F.~Mukhanov, H.~A.~Feldman and R.~H.~Brandenberger,
Phys. Rept. \textbf{215}, 203-333 (1992). 


\bibitem{Cheung:2007st} 
  C.~Cheung, P.~Creminelli, A.~L.~Fitzpatrick, J.~Kaplan and L.~Senatore,
  JHEP {\bf 0803}, 014 (2008). 


\bibitem{Cheung:2007sv} 
  C.~Cheung, A.~L.~Fitzpatrick, J.~Kaplan and L.~Senatore,
  JCAP {\bf 0802}, 021 (2008). 

\bibitem{Akhshik:2015nfa}
M.~Akhshik, H.~Firouzjahi and S.~Jazayeri,
JCAP \textbf{07}, 048 (2015). 

\bibitem{Jarnhus:2007ia}
P.~R.~Jarnhus and M.~S.~Sloth,
JCAP \textbf{02}, 013 (2008). 



\bibitem{Arroja:2008ga}
F.~Arroja and K.~Koyama,
Phys. Rev. D \textbf{77}, 083517 (2008). 

\bibitem{Noumi:2014zqa}
T.~Noumi and M.~Yamaguchi,
[arXiv:1403.6065 [hep-th]].

\bibitem{Weinberg:2005vy}
S.~Weinberg,
Phys. Rev. D \textbf{72}, 043514 (2005)\, .

\bibitem{Domenech:2021ztg}
G.~Dom\`enech,
Universe \textbf{7}, no.11, 398 (2021)
[arXiv:2109.01398 [gr-qc]].


\bibitem{Cai:2018dig}
R.~g.~Cai, S.~Pi and M.~Sasaki,
Phys. Rev. Lett. \textbf{122}, no.20, 201101 (2019). 


\bibitem{Liu:2020oqe}
J.~Liu, Z.~K.~Guo and R.~G.~Cai,
Phys. Rev. D \textbf{101}, no.8, 083535 (2020). 


\bibitem{Talebian:2022cwk}
A.~Talebian, S.~A.~Hosseini Mansoori and H.~Firouzjahi,
Astrophys. J. \textbf{948}, no.1, 48 (2023). 

\bibitem{Ragavendra:2021qdu}
H.~V.~Ragavendra,
Phys. Rev. D \textbf{105}, no.6, 063533 (2022).



\end{thebibliography}
\end{document}